**Title**

Review: dual benefits, compositions, recommended storage, and intake duration of mother's milk

**Authors**

Ammu Prasanna Kumar (APK)[1,2§*], Suryani Lukman (SL)[1§*]

[1]Department of Chemistry, College of Arts and Sciences, Khalifa University of Science and Technology, Abu Dhabi, United Arab Emirates

[2] Research Unit in Bioinformatics, Department of Biochemistry and Microbiology, Rhodes University, South Africa

[§]The authors contributed equally to this study

*Correspondence to: Department of Chemistry, College of Arts and Sciences, Khalifa University of Science and Technology, P.O. Box: 127788, Abu Dhabi, United Arab Emirates,

Emails: suryani.lukman@ku.ac.ae (SL), g19k5181@campus.ru.ac.za (APK)

Telephone: +97124018222, Fax: +97124472442

Short title: Mother’s milk composition, storage and intake.

**Abstract**

Breastfeeding benefits both infants and mothers. Nutrients in mother's milk help protect infants from multiple diseases including infections, cancers, diabetes, gastrointestinal and respiratory diseases. We performed literature mining on 31,496 mother's-milk-related abstracts from PubMed and the results suggest the need for *individualized* mother's milk fortification and *proper* maternal supplementations (e.g. probiotics, vitamin D), because mother's milk compositions (e.g. fatty acids) vary according to maternal diet and responses to infection in mothers and/or infants. We review at details the variability observed in mother's milk compositions and its possible health effects in infants. We also review the effects of storage practices on mother's milk nutrients, recommended durations for mother's milk intake and the associated health benefits.

## Introduction

Mother's milk help infants and preemies to develop immunity[1] and long-term, dose-dependent protections from respiratory infections, gastrointestinal diseases, sudden infant death syndrome (SIDS), childhood cancers (leukemia and lymphoma)[2], obesity, diabetes, multiple sclerosis[3], behavior problems[4], and hand, foot and mouth disease (HFMD) (Supplementary).

In infants born to HIV-infected women, exclusive breastfeeding reduces diarrhea and respiratory infections[5]. Breastfeeding duration also influences the protections against eczema, allergic rhinitis, rhino conjunctivitis, and asthma[6,7]. Even partial breastfeeding can reduce SIDS[8]. Mother's milk may reduce the risk of obesity and diabetes[9]; breastfeeding (for ≥3 months) may reduce overweight in children of gestationally-diabetic women[10]. Mother's milk, especially its longer intake[11], promotes brain growth and cognition[12];

For mothers, breastfeeding reduces their risks for breast cancer[13], endometrial cancer, ovarian cancer[14], obesity-related diseases, postpartum depression and bone loss. Breastfeeding duration is inversely associated with risks for breast cancer[15]. Breast cancer risks are lower for parous women (regardless of birth numbers) who have cumulatively breastfed for 12/> months[16], as breastfeeding alters hormones, molecular histology and immunity[13] through undiscovered mechanisms. Breastfeeding lowers endometrial cancer risk through reduced estrogen[17]. Long-term breastfeeding protects against ovarian cancer[18]. Intense breastfeeding helps with managing pregnancy-gained weight and reducing risks for metabolic syndrome[19], hypertension[20], type 2 diabetes[21] and heart disease[22]. Lactating mothers secrete prolactin and oxytocin that help them to adapt to motherhood[23]. Oxytocin is related to lower anxiety/depression[24]. Oxytocin also helps restoring uterus size and reducing uterine bleeding[25].

Despite these numerous benefits of breastfeeding, globally, less than half of the infants are exclusively breastfed for the initial six months. Studies suggest that knowledge on breastfeeding can improve this rate[26,27]. This may be because mother's who understand the benefits of breastfeeding are more likely to exclusively breastfeed[28]. Although there are many published articles reviewing the benefits of mother's milk, given the importance of educating mothers in promoting breastfeeding, it is important to update the information. For example, more than 6000 PubMed articles have been published during the last five years with the keywords "mother's milk", "human milk", "breast milk", "breastmilk" or "breastfeed*".

This motivated us to review the literature on mother's milk by including results from text mining analysis of PubMed abstracts published till 2019. The results from text mining particularly suggests that ensuring maternal nutrition has a very important role in ensuring infant health during breastfeeding. We also reviewed literature on mother's milk storage, intake duration, associated concerns (Supplementary) and relactation (Supplementary).

## Methods

To provide a state-of-the-art review of the benefits of mother's milk, we retrieved abstracts on mother's milk from PubMed using multiple keywords on March 6th 2019 (Table **1**). Next, we performed text mining on the results from the keyword search that returned the largest number of articles, i.e. 31,496 articles. We created a dataset and a word cloud (Figure **1**) based on the 31, 496 abstracts. The data set had the information on the countries to which the largest number of articles and authors belong to, and the countries that frequently show in article titles. The word cloud gave us clues on to the variability observed in mother's milk. Then we reviewed mother's milk-related publications based on their relevance with respect to the clues obtained from word cloud (see Results), or to the topics- mother's milk storage, intake duration, associated concerns or relactation.

**Table 1**: Number of Articles Retrieved using Different Keywords of Mother's Milk.

| **Keywords Used** | **Number of Articles Retrieved** |
|---|---|
| "mother's milk" | 1327 |
| "human milk" | 21,351 |
| "breast milk" | 12,275 |
| "breastmilk" | 1655 |
| "breastfeed*" | 4398 |
| "breast milk" OR "breastmilk" | 13,773 |
| "mother's milk" OR "human milk" OR "breast milk" OR "breastmilk" OR "breastfeed*" | 31,496 |

## Results and Discussion

### Results from text mining

Several countries were frequently mentioned in article field, including US (13,176), Switzerland (1273), Netherlands (1475 articles), Germany (1499), England (6,385) and authors were predominantly from US (19,329 authors). The article titles frequently referred the countries US (212), Niger (128), Nigeria (123), Kenya (86), Japan (127), India (160), Ethiopia (96), China (156), Bangladesh (87), Brazil (138), Australia (154 times) and the African continent (299). The data indicates that research on mother's milk is highly promoted in US. Interestingly, US has also increased its breastfeeding rate according to their most recent breastfeeding report card, although there is yet to improve especially in exclusive breastfeeding.

The overall quality of mother's milk is highly conserved despite maternal diet[(29)]; it nourishes and protects babies. However, the word cloud suggests that variability is observed in mother's milk composition of macronutrients and micronutrients which can affect infants[(30–32)] (Supplementary). These variabilities are mainly associated with maternal factors including health and diet. For example, less interleukin-13 producing cells had been observed in mother's milk of infants with atopic dermatitis[(33)]; low level of zinc (i) is observed in mother's milk of women with loss-of-function mutation in zinc transporters and (ii) causes transient neonatal zinc deficiency (TNZD) in exclusively breastfed infants[(34)]. The risk of TNZD is at least 1 in 2334 infants[(34)]. High mother's milk cytokine (TGFβ2) amounts is associated with high risks for eczema in infants[(35)]. Mother's milk from women who do not consume food from animal sources may lack some important micronutrients in sufficient quantities such as vitamin B12 and DHA[(36–39)] and these can negatively affect the infant[(37)]. In addition mother's milk also lack sufficient vitamin D[(40)] and this increases infant's risks for rickets, respiratory diseases, type 1 diabetes[(41)], hypocalcemic seizures and dilated cardiomyopathy[(42)]. These studies suggest that early detection of nutrient deficiencies as well as proper supplementation of nutrients are necessary for breastfeeding women.

Timing of nutrient deficiency is important, also in case of the brain development of children[(43,44)]. Some neurodevelopmental processes starts during pregnancy and completes within the first few years of life and nutrient deficiencies in this time period can affect the process[(44)]. The severity of nutritional deficiencies determines their negative effects on infants[44]. Sometimes, the effects may be irreversible when treated later[44–47]. Moreover, nutrients act synergistically causing deficiency of one nutrient to affect another[40,43]. Recovery from the negative effects may only be possible, if the nutrients are available during the time the affected growth process is still occurring[44]. Therefore, improving maternal nutrition in malnourished pregnant and lactating women should be given increasing attention.

The variabilities observed in mother's milk are particularly important in case of preemies since they have higher nutritional requirements and risks for diseases and growth retardation[48]. However, mother's milk is undeniably the best food for the preemies as it offers protection and ensures their proper development[48]. Therefore, individualized mother's milk fortification[49] can be used to address this variability.

Next, we discuss (i) the mother's milk compositions and the variability in greater details, (ii) effects of storage on expressed mother's milk, (iii) recommended durations for breastfeeding and factors affecting the duration. See Supplementary for discussion on (i) concerns associated with mother's milk and (ii) relactation.

## Mother's milk compositions and their variability

To meet infant developmental needs, mother's milk changes from colostrum, transitional milk, to mature milk (Supplementary) with distinct compositions. Mother's milk nutrients (proteins, fats, carbohydrates, vitamins) can promote infant health and survival[29] (Supplementary). Mature milk contains 60-75 kcal/100 ml and its fat, proteins and carbohydrate contents are ~3-5%, 0.8-0.9% and 6.9-7.2%, respectively[50]. During early lactation, mother's milk contains more whey (protein that remains as liquid, e.g. lactoferrin, alpha-lactalbumin, immunoglobulins, lysozyme) than casein (protein that becomes curds in stomach and harder to digest)[51,52]. Subsequently, their proportions equalize to facilitate nutrient absorption and immunity in infants[52]. Mother's milk provides essential fatty acids (linoleic and alpha-linolenic acids) for infant growth and brain development[53]. Mother's milk fats (>98% triglycerides of palmitic and oleic acids) increase throughout a single nursing[54]. More than 98% of the fat is in the form of triglycerides[55]. Main mother's milk carbohydrates include lactose, that induces innate immunity by triggering antimicrobial peptides production[56], and oligosaccharides, that shape microbiota and protect against Group B *Streptococcus* infection[57]. Mother's milk completely provides carotenoids[58] for immunity and vision.

Although mother's milk has all essential fatty acids, its DHA level is variable and depends on maternal diet, particularly on fish products that are rich in preformed DHA[37,59]. The rate of synthesis of DHA from alpha-linolenic acid, its precursor, is also inefficient for infants and the DHA level depends on maternal supply through mother's milk[37]. DHA is essential for neurovisual development; its deficiency can negatively affect mental development and learning abilities[37]. Maternal diets also influence vitamin composition in mother's milk, especially the composition of vitamins A, B1, B2, B3, B6, B12, C and D[60–66]. Deficiency of vitamin B6, vitamin B12, and thiamine causes growth stunting[61]. Mother's milk minerals are important for infant growth, development and immunity[67]. While mother's milk calcium, magnesium, iron, zinc, copper amounts are independent of maternal mineral status[68–70], selenium amount is dependent[71].

Since maternal nutrients affect mother's milk compositions, maternal supplementations are available. While folic acid intake is recommended in the first trimester to avoid fetal neural tube defects, it can alter gene expression epigenetically[72] and cause respiratory allergies in mice[73]. Studies of the association between maternal folic acid intake beyond the first trimester and childhood allergic diseases remain inconclusive[72]. While some studies reported their positive association[74,75], maternal intake of folic acid might promote cow's milk allergy and childhood immunoglobulin E antibody-mediated allergic diseases[72,76]. In contrast, pregnancy supplementations of omega-3 polyunsaturated fatty acids (i.e. EPA and DHA) protect against childhood allergies[77].

Mother's milk anti-inflammatory cytokines offer passive protection to infants from intestinal inflammation[78] whereas pro-inflammatory cytokines stimulate active immunity[79]. TGFβ, the dominant mother's milk cytokine[29], promotes mucosal immunity by promoting immunoglobulin A production and may decrease allergic risk[80]. Colostrum has the most leukocytes[81], that significantly increase upon infection.

In addition to maternal diet, maternal lifestyle, geographic environment and lactation time also affect mother's milk composition[77,82–84] and infant health. More research is required to probe the outcomes of diverse mother's milk composition.

## Expressed milk storage

More mothers are expressing milk while being away from their baby due to illness, premature delivery, or returning to work. Expressing milk also relieves engorgement. To prevent microbes' growth, expressed mother's milk (if not for immediate consumption) needs cooling/freezing. Storage temperature, freezing, thawing, light exposure, pasteurization and storage container affect mother's milk composition/quality (Supplementary)[85,86].

Creamatocrit and human milk analyzer (HMA) are used for measuring mother's milk fat and caloric contents. A study on frozen (–20°C) mother's milk samples for 28 days, that used the creamatocrit method, reported that caloric contents did not change much[87]. However, in a study that used the HMA method, Garcia-Lara et al. observed that frozen (at less than –20°C) mother's milk contains less fat and calories than fresh mother's milk[88]. They observed that fat and caloric contents decrease as duration (7, 15, 30, 60, and 90 days) of freezing increases. In their study, fresh mother's milk had 4.52-5.64 g/100 ml of fat and 0.72-0.82 kcal/ml of calories, whereas frozen mother's milk had only 3.84-5.14 g/100 ml of fat and 0.65-0.77 kcal/ml of calories after 90 days of freezing. Freezing for 7, 15, 30, 60, and 90 days, did not significantly change the nitrogen content in mother's milk, and the decrease of lactose in frozen mother's milk was only significant after 90 days of freezing. These results are important because mother's milk is often frozen at less than –20°C in neonatal units and human milk banks[88]. Freezing at less than –80°C is better yet much costlier[88].

The inconsistencies in the results of the above mentioned studies may be due to the difference in the parameters used for energy measurement. Creamatocrit estimates energy based on percentage of cream in mother's milk. For example, to account for the effects of freezing on fat contents in mother's milk using creamatocrit, Wang *et al*. used the following regression equations to calculate caloric contents in fresh and frozen mother's milk: for fresh mother's milk, energy (kcal/dl) = 5.99 x creamatocrit(%) + 32.5; for frozen mother's milk, energy (kcal/dl) = 6.20 x creamatocrit(%) + 35.1[89]. On the other hand, in HMA, protein, lactose and fat in mother's milk can be measured[90]. In the study of Garcia-Lara et al.[88], caloric contents in mother's milk was calculated using this equation: energy (kcal/dl) = 9.25 × fat + 4.40 × total nitrogen + 3.95 × lactose[88]. While several studies suggest that creamatocrit correlated strongly with fat and energy content in mother's milk[89,91,92], Neill et al. reported that creamatocrit overestimates fat and energy content in mother's milk and HMA has increased accuracy[90].

Mother's milk fat and protein content are reduced by pasteurization[93]. Fat content is significantly reduced while delivery using continuous infusion feeding, due to the adherence of fat to the delivery system[93]. Protein denaturation occurs during freezing and thawing[94]. Freezing mother's milk can change and destabilize casein micelles and change the quaternary structure of whey proteins, resulting in the formation of precipitates[88]. However, storage temperatures have not shown to influence the protein concentration in mother's milk[94]. Reduction in fat and protein concentration is a serious concern as it can affect the growth rate of preterm infants[95].

Milk fat globule membranes (MFGMs) in mother’s milk are broken by freeze, thawing procedures[88,96]. The triglycerides released with rupturing of MFGMs can come in contact with the lipase enzymes in mother’s milk[96], which are observed to be active at freezing temperature (–20°C), and room temperatures (5°C, 25°C and 38°C)[85,88,96]. The lipase catalyzes the hydrolysis of triglycerides (also known as lipolysis) into diglycerides, monoglycerides, and free fatty acids[97]. Triglycerides are esters derived from glycerol and three fatty acids. In a diglyceride, its glycerol is ester-linked to two fatty acids, whereas in a monoglyceride, its glycerol is ester-linked to only a fatty acid. The fatty acids produced by this hydrolysis have antimicrobial activity[98–100]. On the other hand, rupturing of MFGM may reduce the bactericidal properties of mother's milk as MFGMs have bacterial sequestration abilities[96]. However, at –20°C, a significant reduction in bactericidal activity of mother’s milk is not observed after one month[96]. Freezing at –70°C should be optimal as fat hydrolysis does not occur at this temperature[101].

Another matter of concern is lipid peroxidation. Lipid peroxidation is the process in which free radicals (most commonly reactive oxygen species) remove electrons from lipids producing reactive intermediates[102]. Turoli et al. observed lipid peroxidation in fresh and frozen mother's milk (–20°C), and formula milk[103]. They observed highest lipid peroxidation in frozen mother's milk. This finding may be related to the higher contents of free fatty acids in frozen mother's milk due to hydrolysis activity of lipase. Lipid peroxidation is also observed in mother’s milk at 4°C[104] and –80°C[105]. However, lipid peroxidation is minimal in frozen mother's milk (–20°C and –80°C) compared to that stored in refrigerator (4°C)[104,105]. To minimize lipid peroxidation, freezing temperature of –80°C is preferable over –20°C[105]. Presence of lipid peroxidation products in infant feeds may play roles in development of diseases including necrotizing enterocolitis and bronchopulmonary dysplasia[106]. Studies to determine the effect of consuming lipid peroxides in infants including preemies are needed.

The antioxidant activity of mother's milk, which could protect against the effects of lipid peroxidation, also decreases with storage at refrigeration (4°C) and freezing temperatures (–8°C, –20°C, –80°C)[105,107,108]. The complete list of antioxidant components in mother's milk and their contribution to this variation are unknown. However, the carotenoid (major antioxidants in mother's milk) contents in frozen mother's milk were reported to be stable at -18°C for 28 days[109]. Storage at –80°C for a period of less than 30 days is recommended to maximally preserve antioxidant activity[105].

Freezing (–20°C/–70°C) and heat treatment have little effects on concentration of vitamins A, D and E in mother's milk[107,110,111]. On the other hand, storage for more than one month in a freezer (–16°C) or 24 hours in a refrigerator (4°C-6°C) may substantially reduce vitamin C concentration[112]. Holder pasteurization may affect the vitamin B6 concentration in mother's milk[110].

Many protective immunologic components of frozen mother's milk kept at –20°C remain stable for a duration of up to 365 days while lactoferrin and live cells amounts are affected by storage container, freezing and/or thawing[113]. Mother's milk lactoferrin significantly decreases after 3 months of freezing (–20°C)[114]. Cycles of freezing and thawing can rupture cells in mother's milk [115] and longer storage time reduces the cell function[113]. Frozen mother's milk kept at –20°C has lower number of cells and functions of surviving cells than fresh mother's milk does[113]. Cells in mother's milk can also adhere to the surface of the storage container particularly Pyrex glass container[86].

**Duration of mother's milk intake**

In an Australian study, only 45.9% of babies received any mother's milk at age 6 months; 19.2% at age 12 months[116]. Maternal attitudes are among factors lengthening the duration of mother's milk intake[116–118]. Opposing factors include early breastfeeding difficulties[116], inadequate mother's milk[119], lack of family support[120], early introduction of pacifier[116], smoking[116], social stigma[121] and short maternity leave[116]. Since duration and exclusivity of mother's milk intake are influencing health benefits (Supplementary), lactation counseling can promote breastfeeding exclusivity and its total duration[122]. A USA study comparing exclusive mother's milk intake for 6 months/more versus 4-<6 months, found that the former has lower respiratory tract infections[115]. Breastfeeding beyond 6 months can better protect against respiratory and gastrointestinal tract infections[123].

Besides studies supporting World Health Organization (WHO)'s recommendations of *exclusive* breastfeeding for the first six months[124,125] and continued breastfeeding up to 2 years/beyond along with complementary foods, some studies suggested that exclusive breastfeeding may be insufficient for 6-month-old infants[126,127]. However, early introduction of complementary foods may negatively affect infants[128,129]. Prolonged breastfeeding after 6 months along with complementary foods is beneficial for mother-child bonding[130], healthy dietary patterns at age 6[131], and decreasing asthma risk[6].

After infants turn 6 months old, complementary foods are essential, because zinc, calcium, vitamin B6 and vitamin C levels in mother's milk subsequently decrease[132,133], whereas mother's milk continues providing vitamin A (first 3 years)[134,135] and fat (first 2 years)[134,136]. In the second lactating year, mother's milk provides significantly more protein, lactoferrin, immunoglobulin A and lysozyme[137].

Considering mother's milk benefits, relactation may be necessary after mother-infant separations due to e.g. health issues. Relactation efforts (Supplementary) deploying physiology-based methods and galactagogues (including fenugreek and prescription drugs domperidone and metoclopramide with side effects[138]) have shown successes. The former methods promote prolactin and oxytocin releases[139].

## Conclusion

Mother's milk protects against various diseases in infancy, childhood and later life, by promoting proper and long-term immunity. More future studies are needed to understand the molecular mechanisms of how mother's milk influences maternal hormonal levels and renders immunity to infants. Further studies are also necessary to know more on variabilities observed in mother's milk. A strategy that builds on our current work, will be exhaustively mining the literature on mother's milk and hormones including estrogen, prolactin, and oxytocin. Infants, including infected ones, require mother's milk. Knowledge of mother's milk compositions, storage and intake duration can help mothers to make better efforts in breastfeeding and lactation.

**Conflict of Interest**

None.

**Acknowledgment**

We thank our colleagues at Khalifa University for suggestions and proofreading our manuscript, particularly Jessica Allred (English lecturer).

**Figure 1**

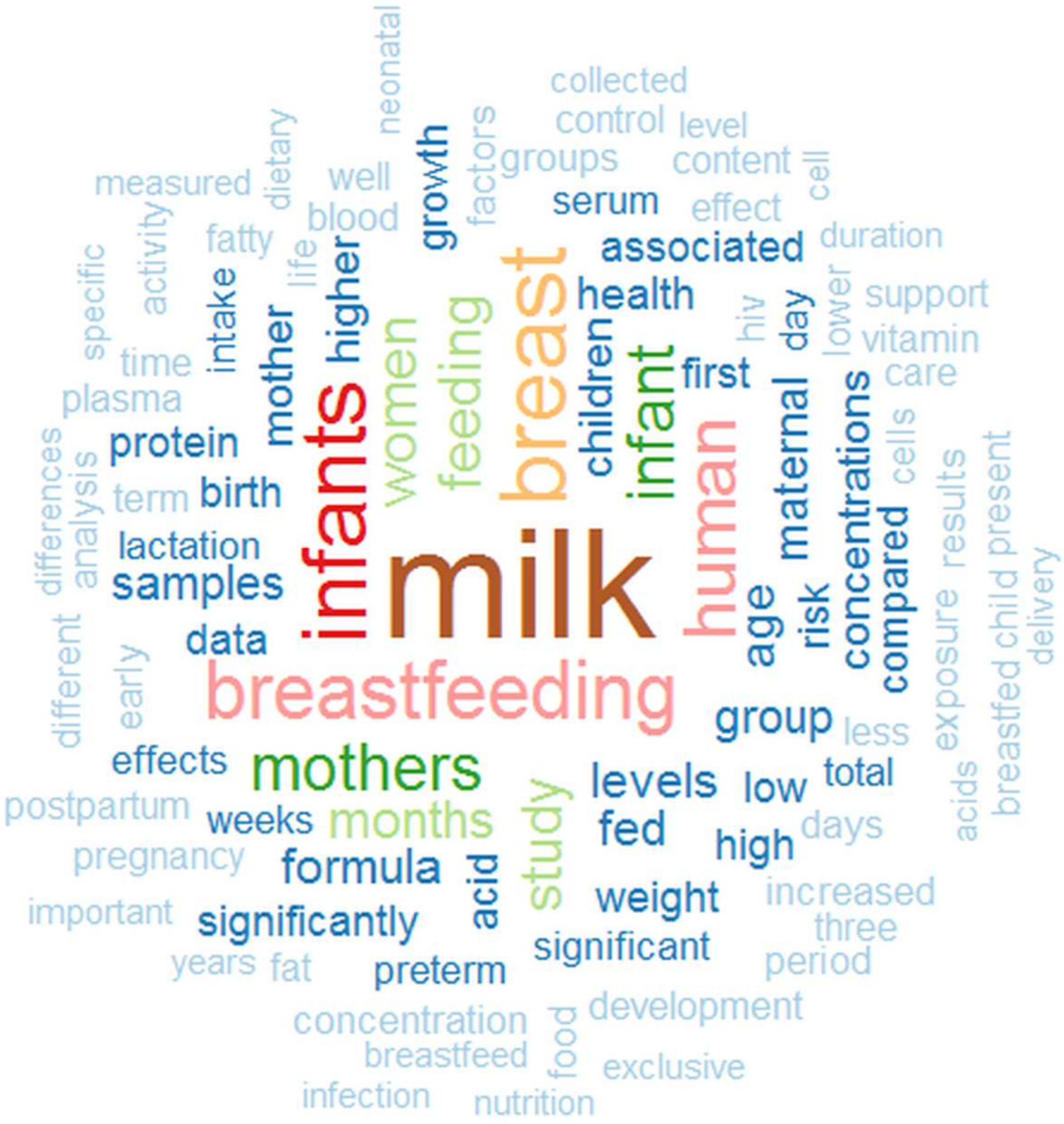

## Figure Legend

**Figure 1: Text mining of mother's milk and breastfeeding literature.** The word cloud shows the most frequent words in 31,496 abstracts retrieved from PubMed using the keywords "mother's milk" OR "human milk" OR "breast milk" OR "breastmilk" OR "breastfeed*".

## Supplementary data

**Contents and Page Numbers**

**Mother's milk classifications**

Breastfeeding offers numerous benefits for the mother and the child[(1)]. From the time an infant is born, mother's milk undergoes changes in its composition, to meet varying developmental needs. Sequentially, mother's milk can be classified as colostrum, transitional milk and mature milk. Colostrum (the initial milk produced after giving birth) is thick, yellowish in color, low in volume, and provides the complete nutrient and immunological protection for the infant[(2–4)]. After 4-5 days, transition milk replaces colostrum. Transition milk is creamy, has higher volume and its production may last up to two weeks[(3)]. Subsequently, mature milk is thinner, watery and has greater volume than transition milk[(5)].

**Benefits of mother's milk consumption for infants**

Besides providing all the essential nutrients to infants, mother's milk also helps infants to absorb nutrients, develop immune system and fight infections[(6,7)]. Breastfed babies gain protections from infections including respiratory infections, gastrointestinal tract diseases, hand, foot and mouth disease, atopic diseases, sudden infant death syndrome (SIDS), childhood cancers, obesity, diabetes, multiple sclerosis and childhood behavior problems. Mother's milk also promotes brain development[(8–11)]. Many of these effects are observed long term and are dose dependent. Breastfed infants have lesser risks for hospital stay and morbidity due to various diseases[(12,13)]. Breastfeeding is also highly beneficial for premature babies (born before 37 weeks of gestation), who are at higher risks for infections and cognitive impairments[(14–17)]. In addition, mother's milk is also friendlier to the tummy of the sucklings as it can be easily digested in their stomach[(6)]. Breastfeeding promotes esophageal peristalsis and reduces the duration of gastro esophageal reflux (a condition where stomach contents flow backward to the esophagus resulting in heartburn, chest pain, vomiting etc) in infants[(18,19)].

Breastfeeding offers long term protection against respiratory infections, gastrointestinal tract diseases, hand, foot and mouth disease, dose dependently[(20–28)]. Tromp *et al.* observed that breastfeeding was associated with reduced risks of lower respiratory tract infections up to 4 years of age[(24)]. They observed that the association was dose dependent and the results were not significant for breastfeeding for <6 months. In agreement with this observation, Wang *et al.* observed that breastfeeding for >6 months protects against bronchiolitis (a lower respiratory tract infection) up to two years of age[(25)]. Duijts *et al.* observed that exclusive breastfeeding for 6 months have a strong protective effect against upper respiratory tract, lower respiratory tract and gastrointestinal tract infections until 6 months of age and lower respiratory tract infections until 12 months of age[29]. Li *et al.* have shown that breastfeeding may provide long term protection against ear, throat and sinus infections until 6 years of age depending on duration and exclusivity of breastfeeding[28]. In their study, they observed that children who were exclusively breastfed ≥6 months had lower experience of these infections. Mwiru *et al.*, have shown that exclusive breastfeeding is associated with reduced risks for diarrheal and respiratory infections in infants born to the HIV-infected women[30]. In another study, Kramer *et al.* suggested that exclusive breastfeeding, for the first six months, may compromise for iron deficiencies and reduce the risks for gastrointestinal tract infection in infants[31]. Breastfeeding has been suggested to protect against inflammatory bowel diseases such as Crohn's disease and ulcerative colitis[32]. Exclusive breastfeeding for the first 6 months has also been suggested to have a protective effect against hand, foot and mouth disease among children upto 28 months of age[26].

Odijk *et al.* observed that breastfeeding appears to protect against the development of atopic diseases (including eczema, asthma[33], allergic rhinitis, and rhinoconjunctivitis), depending on duration of breastfeeding[34]. In children with hereditary allergies, the beneficial effects of mother's milk are more observed [34]. The protective effects of breastfeeding increases with its duration. They observed that, when mother's milk is insufficient, extensively hydrolyzed cow's milk formula reduces the risk for eczema and asthma[34]. The protective effects of breastfeeding against asthma and eczema are more apparent in middle- and low-income countries[33].

Breastfeeding can reduce the risk of SIDS throughout infancy[35,36]. More recently, Thompson *et al.* have found that both partial and exclusive breastfeeding for at least two months is associated with reduced risk for SIDS, with greater protection with increased duration[37]. They found that breastfeeding does not need to be exclusive to give this protection.

Breastfeeding is associated with reduced risk of childhood cancers involving leukemia and lymphoma[38–41]. There are inconsistencies regarding the association of duration of breastfeeding and the risk of childhood cancer[40,42].

Breastfeeding may reduce the risk of obesity, type 1 and type 2 diabetes in later life[43–48]. Shields *et al.* observed an association between shorter durations of breastfeeding and increased risks for overweight at 14 years of age based on a study involving 3698 children[49].

Breastfeeding for at least three months may reduce the risks for overweight in children of gestationally diabetic women[50]. Longer duration of breastfeeding can be especially beneficial to lower the risk for diabetes, particularly type 1 diabetes[51,52].

Longer duration of breastfeeding is associated with reduced risks for multiple sclerosis (an immune-mediated disease that affects brain)[53,54]. Liu *et al.* have shown that breastfeeding and active interaction of the mother with the infant can protect against childhood internalizing behavior problems[55].

Breastfeeding promotes brain development and improved cognitive skills[8–11]. In a study involving 300 participants, infants who consumed mother's milk in the early weeks of life had a significantly higher intelligence quotient (IQ) at 7.5-8 years than those who received no mother's milk[8]. Breastfeeding is associated with larger whole brain and improved development of white matter (in later maturing frontal and association brain regions)[9], total gray matter, total cortical gray matter and subcortical gray matter[56] and these impacts on brain structure possibly leads to the higher IQ observed in breastfed children. Longer duration of breastfeeding positively influences this brain growth[57] and the positive effects can even be observed beyond infancy; Kafouri *et al.* have shown that prolonged exclusive breastfeeding is positively associated with thicker parietal cortex in adolescents[10].

Mother's milk also protects preemies. Mother's milk can offer protection against infection and sepsis/meningitis in preemies[58–61]. Increased intake of mother's milk has been reported to protect the infant against necrotizing enterocolitis (NEC) (a disease in which bacteria invade the intestinal walls of premature infants)[14,15]. Breastfeeding also promotes brain growth in preterm infants[16,17]. In a study involving infants born at <30 weeks' gestation or <1250 grams birth weight, consuming over 50% of mother's milk as daily intake in the first four weeks of life, seemed to associate with larger deep nuclear gray matter volume and higher cognitive functions (including IQ), academic achievement, working memory, and motor function at age 7[62]. Mother's milk has also shown to reduce risk for developmental and behavior morbidities in very low birth weight infants[63].

In summary, breastfeeding protects against various diseases in infancy, childhood and later life. Studies suggest two possible reasons for the protective effects of mother's milk in later life[28]. One is that breastfeeding promotes proper development of the immune system in infants, which protects them from development of diseases even in later life. The other is that the long term protection may be mediated through protection from illnesses during infancy, which are the risk factors for other diseases later in life. However, several findings that are inconsistent with the above mentioned observations on the benefits of breastfeeding have also been reported[33,64]. For example, Lodge et al. observed that there is weaker evidence for the protective effect of breastfeeding against allergic diseases such as asthma, eczema and allergic rhinitis[33]. They observed that these protective effects are greater in early life and are more apparent in middle-/low- income countries. They suggested the protective effect of breast feeding against respiratory infections, one of the risk factors for asthma, may be offering this protection. Additionally, since many of the evidences on the benefits of breastfeeding are provided by observational studies, it is possible for them to be biased by various factors including methodological quality, heterogeneity in study design and characteristics of people involved in the studies[33].

**Benefits of mother's milk production for mothers**

Breastfeeding reduces the mother's risks for various cancers including breast cancer [65–68], endometrial cancer and ovarian cancer[69], obesity related diseases, postpartum depression and bone loss.

Breastfeeding can reduce the risks for basal-like breast cancer[70,71], triple-negative breast cancer[72] and luminal B breast cancer[73]. Several studies suggest that duration of breastfeeding is inversely associated with risks for breast cancer[66]. Anothaisintawee et al., suggested that parous women have lower risks for breast cancer, irrespective of the number of births and when the cumulative duration of breastfeeding was 12 months or longer, the risks are lower[74]. Studies suggested that alterations in hormones, molecular histology and immune responses associated with breastfeeding provides this protective effect against breast cancer[65,68]. However, the exact mechanisms remain unknown.

High estrogen levels can increase the risks for endometrial cancer. Breastfeeding is associated with reduced estrogen levels, and studies show that breastfeeding can reduce the risks for endometrial cancer[75,76]. Rosenblatt et al., suggested that even though prolonged lactation can reduce the risks for endometrial cancer, the effect may not persist at later life[77].

Breastfeeding has inverse association with risks for ovarian cancer, especially long-term breastfeeding has been suggested to provide a stronger protection against ovarian cancer[78–80)]. Feng et al., suggested that breastfeeding for 8-10 months may be effective to provide protection against ovarian cancer[81].

Fat stores that accumulates in women's body during pregnancy leads to weight gain[82] and is a risk factor for obesity and related diseases[83–86]. Intense breastfeeding can rapidly mobilize these fat stores[87], help in weight management[88,89] and reduce the risks for the associated diseases including metabolic syndrome[90], hypertension[91], type 2 diabetes[92] and coronary heart disease[65,93,94].

Prolactin and oxytocin, two hormones secreted while breastfeeding[6], help the women physiologically and behaviourally adapt to motherhood[95–98]. In a study involving women intending to breastfeed, Steube et al. have observed association between higher depression and anxiety with lower levels of oxytocin[99]. Oxytocin also helps restoring uterus size and reducing uterine bleeding[100].

During pregnancy, a growing baby's increased need for calcium may alter calcium homeostasis and may lead to bone loss in the mother[101]. However, lactation following pregnancy has shown to have a protective effect on maternal bone health[102]. Since evidences for the mother's milk's benefits are provided by observational studies *hitherto*, they may be biased by methodological quality and heterogeneity in study design [33].

**Concerns of mother's milk**

We discuss here the various concerns associated with mother's milk such as the possibility of transfer of diseases, medications and contaminants from the mother to the infant, effects of mother's smoking and alcohol consumption on infant, adequacy of mother's nutrition, and food restrictions during breastfeeding.

While infected mothers excrete causative agents to their milk[103], for most of the diseases, the benefits of mother's milk for infants outweigh the associated risks, and cessation from breastfeeding is not recommended. The infant is already exposed to the infection at the time when the initial symptoms of the disease appears in the mother, therefore denying the mother's milk for infant may deprive him/her of the protection offered by the mother's milk[104]. However, for infections caused by RNA retroviruses such as HIV, HTLV-1 and HTLV-2, breastfeeding should be combined with medications for the mother and the infant as the infections can transfer to and affect the infant through mother's milk[103,105,106]. Evidence suggest that mothers with cytomegalovirus (CMV) infection also transfer the pathogenic agents via mother's milk to the infants[106]. However, maternally acquired antibodies from the placenta protect the infants from CMV infection. On the other hand, preterm infants who may lack these antibodies have higher risks for this infection[104]. Although Zika virus is also detected in mother's milk[107,108], no adverse effects of the virus on infants through milk consumption of infected mothers have been documented[109]. It is also possible for the milk to be contaminated by blood, for example through bleeding nipples of infectious mothers[110]. Maternal infectious diseases and the associated breastfeeding recommendations are reviewed in detail elsewhere[104,106].

Many women are advised to stop breastfeeding while taking any medication. Such a concern is unnecessary since most of the drugs transfer only in small amount via mother's milk and are unable to cause any adverse effects in infants[111,112]. However, there are few drugs that are of concern. Some drugs are known to interfere with infant metabolism and cause adverse effects in infants and these includes several cytotoxic drugs, drugs of abuse (including cocaine, heroin and marijuana), anti-anxiety drugs, pain medications (including narcotic agents such as codeine and hydrocodone) and radioactive compounds[111,113]. Information on various drug levels in mother's milk, their adverse effects on infants and possible alternatives can be found in https://toxnet.nlm.nih.gov/newtoxnet/lactmed.htm. Taking medications just after breastfeeding may minimize the drug exposure of infants[113].

Several chemical contaminants and potentially toxic metals have been detected in mother's milk in the past years. The chemicals include DDT, hexachlorobenzene, cyclodiene pesticides, and polybrominated diphenyl ethers[114]. The metals include lead and mercury[114]. These contaminants that accumulates in mother’s body from environment exposure or food transfer via mother's milk to the infant[114]. Women with high body mass index are at higher risks for the accumulation of lipophilic chemicals, which accumulates in the adipose tissue[114]. However, breastfeeding benefits outweigh the risks associated with the intake of these contaminants[114]; The risks can also be minimized through proper diet and weight management[114].

Maternal smoking increases the risk of infant’s smoking in adulthood[115]. Mothers who smoke transfer nicotine to mother's milk[116]. Even though breastfeeding can offer protection against some adverse effects of nicotine on infant health[117], early exposure to nicotine may make it appealing for infants in later life[118] and causes sleep disruption in infants[119].

Mennella et al. suggested that sleep disruption during infancy may produce long term behavioral and learning deficits[119]. Parental smoking is also suggested to be an important risk factor for SIDS[120]. However, it is unknown whether some other chemicals from cigarette smoke is transferred via mother's milk to the infant.

Mother's alcohol consumption affects both the mother and the infant. Alcohol can transfer through mother's milk to the infant[121]. Alcohol exposure can cause sleep disruption in infants and thus may lead to learning problems[122]. Even in small quantities, alcohol may affect the release of hormones that promote milk production and thus affect the milk supply in the mother[121,123].

Mother's milk nutrients come from either the nutrient reserves of the mother or her diet[124]. Studies suggest that maternal nutrition can affect the mother's milk composition of several vitamins, fatty acids and minerals[125–131]. For example, concentration of vitamins A and D in mother's milk decreases, when mother's diet is deficient of these nutrients[124], suggesting the importance of ensuring good nutrition to the breastfeeding women. On the other hand, there are no foods to be avoided during breastfeeding, unless an infant reacts negatively to it[132]. For example some women avoid caffeine and spicy foods during breastfeeding in the fear that the infant may be negatively affected[132]. Such diet restrictions may cause discomfort to a mother and discourages her from continuing breastfeeding[132]. Another myth is that drinking excessive fluids helps in increasing mother's milk volume. However, increase in fluid intake did not appear to increase mother's milk volume[132,133].

In summary, previous studies give the following suggestions- (i) for most of the maternal diseases, breastfeeding can be continued, (ii) most of the maternal medications do not pose any health hazards to the infant, (iii) breastfeeding benefits outweigh the risks associated with the presence of contaminants in mother's milk, (iv) maternal smoking and alcohol consumption may affect the infant, (v) breastfeeding mothers need good nutrition and (vi) food restrictions during breastfeeding may be unnecessary.

**Compositions of mother's milk**

An example of mother's milk proteins, lactoferrin can fight against pathogens[134], facilitate iron absorption into cells[135] and provide immunity[136]. Secretory immunoglobulin A (IgA), which is the most abundant immunoglobulin found in mother's milk, can promote the development of infant's mucosal immune system, fight against pathogens and provide immunity[136]. Studies show that preterm milk (milk produced by mothers on preterm delivery) and colostrum have significantly higher amount of proteins[137,138].

The fat content in mother's milk increases from the beginning to the end of a single nursing[139]. More than 98% of fat in mother's milk is in the form of triglycerides[140]. Triglycerides are esters derived from glycerol and three fatty acids. Palmitic and oleic acids form the major proportion of fatty acids in mother's milk triglycerides[3]. They are heavily concentrated in the 2-position and in the 1- and 3- positions of the triglycerides, respectively[141]. Mother'smilk is also an important source of the essential fatty acids such as linoleic acid and alpha-linolenic acid, which are substrates for the synthesis of the fatty acids- arachidonic acid (AA) and Docosahexaenoic acid (DHA), that are essential for growth, brain development and health of infants[142–144]. The proportions of the essential fatty acids in mother's milk is affected by maternal diet [129–131,145,146].

The principal carbohydrates in mother's milk are lactose and oligosaccharide [3]. Cederlund et al. have shown that lactose can stimulate the expression of cathelicidin antimicrobial peptide (CAMP) gene and its product hCAP-18 , that plays essential roles in fighting against infections and shaping microbiota[147]. Thus lactose is suggested to have a role in inducing innate immunity in newborns[147]. Oligosaccharides also act as prebiotics, facilitating the growth of intestinal microbiota, thus protecting the infant from infections[148,149]. Recent evidence suggest that oligosaccharides may protect against Group B Streptococcus infection[150]. Coppa et al., in a study of 46 mothers, observed increase in levels of lactose and decrease in levels of oligosaccharides in mother's milk over the first 4 months after delivery[151].

Vitamin composition in mother's milk is influenced by maternal status of the nutrients[3,125,142]. The concentration of many vitamins in mother's milk decreases with the poor maternal status of the nutrients[125]. Studies show that the diet and body stores of breastfeeding and lactating mothers are important factors affecting mother's milk composition of vitamins, especially the composition of vitamins A, B1, B2, B3, B6 B12 and D[124–128]. When the vitamin composition of mother's milk is not affected by maternal diet, their low intake may cause their deficiency in mothers. For example, with low folate intake, the mother may become depleted of this nutrient even though its concentration in mother's milk will be maintained[125]. Deficiency of nutrients can affect the health and development of infants. For example, deficiency of vitamin B6, vitamin B12, and thiamine are known to cause growth stunting in infants[125]. However, future research is needed to understand the potential effects of maternal nutrient deficiency on infants.

Mother's milk is also a complete source of carotenoids to the infant until 6 months of age[152]. Carotenoids are essential for immune function, vision, protecting against degenerative eye diseases[153]. Human body cannot synthesize carotenoids, but acquire them from consuming vegetables, particularly those of yellow, orange, and red in colors. Some carotenoids (e.g. β–carotene) can be converted to vitamin A, they are termed as provitamin A carotenoids. In the first year of postpartum time, retinol (preformed vitamin A) content in mother's milk decreases as the postpartum duration increases[154]. Lactating mothers of lower parity and malnourished produce milk with lower content of retinol than those of higher parity and well-nourished [154]. However, mother's milk carotenoids particularly β–carotene can also supply retinol to the infant[155]. Based on a study on 509 mothers in China, Xue et al. suggested that carotenoid composition in mother's milk gradually decreases and reaches a stable level after 12 days postpartum[156]. Several other studies suggest that carotenoid contents stabilizes at nearly 4 weeks postpartum[152,157].

Mother's milk has a variety of minerals having important functions such as supporting growth, development and immunity of the infant[158,159]. Previous studies suggest that the concentration of minerals such as calcium, magnesium, iron, zinc and copper in mother's milk does not depend upon the maternal mineral status[160–162]. On the other hand, the concentration of selenium is influenced by maternal selenium status[161,163]. However, inadequate intake of minerals may lead to maternal deficiency of the nutrients. For example, inadequate maternal intake of iodine causes maternal iodine depletion, though the infant may not be affected[164,165].

Since maternal status of nutrients can affect some of the mother's milk nutrients, maternal supplementation of the nutrients can be considered as a solution. However, studies show that maternal supplementation of nutrients can have both adverse or beneficial effects. For example, to decrease the risk of neural tube defects in fetuses, folic acid has been recommended for pregnant women during the first trimester. Since folic acid is a methyl donor, it can potentially alter gene expression epigenetically[166]. In mice, maternal folic acid supplementation has shown to promote allergic airway disease in the offspring[167]. Several studies have attempted to probe the association between maternal folic acid intake beyond the first trimester and childhood allergic diseases[168,169] and their results are inconsistent[166]. It is important to note that some studies reported their positive association[166,170,171]. Maternal intake of folic acid has been suggested to promote cow's milk allergy[172] and childhood IgE antibody-mediated allergic diseases. On the other hand supplementation of omega-3 polyunsaturated fatty acids (i.e. EPA and DHA) during pregnancy have shown to protect against childhood allergies[173].

Besides the nutrients mentioned above, mother's milk includes several other bioactive components such as growth factors and immunological factors that can promote the health of infants[3,174]. Growth factors include brain-derived neurotrophic factor (BDNF), epidermal growth factor (EGF), erythropoietin, vascular endothelial growth factor, platelet-derived growth factor, insulin and hepatocyte growth factor. These growth factors appear to play roles in promoting health of infant's intestine, vascular system, enteral nervous system and metabolism[3].

Mother's milk contains several anti-inflammatory and pro-inflammatory cytokines that can protect the infant's intestine and provide immunity[175]. Through the anti-inflammatory cytokines, mother's milk offers passive protection to the infant from diseases including intestinal inflammation [176] and the pro-inflammatory cytokines helps in stimulating the immune functions in infants, thereby promoting infant's active immunity.

Transforming growth factor beta (TGFβ) is the most prominent cytokine in mother's milk[3]. Mammals express three different isoforms of TGFβ, they are TGFβ1, TGFβ2, and TGFβ3. Mother's milk contains both TGFβ1 and a majority of TGFβ2 isoforms, at both mRNA and protein levels[177]. In infants, maternal TGFβ promotes mucosal immunity, inhibits pro-inflammatory cytokine release and stimulates the production of endogenous TGFβ[178,179]. The roles of TGFβ in mucosal immune system involves promoting the production of IgA, which improves mucosal defense[180] and induction of oral tolerance[177]. Higher levels of TGFβ in mother's milk[181], particularly colostrum[177], seem to be associated with a decreased risk of allergy. The higher levels of TGFβ in colostrum may compromise for the lower endogenous TGFβ in infant's immune system[177]. Mature mother's milk of allergic mothers contains significantly less TGFβ1 than that of non-allergic mothers[182]. Lower levels of TGFβ1 is associated with necrotizing enterocolitis (NEC) in preterm infants[183]. Cytokines in mother's milk are reviewed in detail elsewhere[175,178].

Mother's milk is also a source of live cells and diverse microbiome[184,185]. Some of these live cells originates from breast (includes lactocytes, mammary stem cells and epithelial cells) and some others from blood (includes hematopoietic stem cells and leukocytes)[184] of the mother. Leukocytes provides active immunity to the infant by promoting immunity and fighting against pathogens[184,186]. Hassiotou et al. have shown that the levels of leukocytes are highest in colostrum and decreases significantly in mature milk[185]. They found that upon maternal and/or infant infection, its level significantly increases in the mother's milk. However, the exact roles of the other live cells in infant are not known. The probiotics present in mother's milk may protect the infant from infection by more severe pathogens[184,187]. Several studies suggest that maternal supplementation of probiotics during lactation, can affect the levels of immunological factors present in mother's milk and may possibly protect the infant from autoimmune, allergic and metabolic diseases[188,189]. However, more studies are needed to confirm the benefits of the maternal supplementation for infants.

Variability observed in mother’s milk macronutrient variability can affect the growth pattern in infants. For example, exclusive breastfeeding is inversely associated with infant weight gain in the first year of life as reported by a Canadian childbirth cohort study recently[190], some other studies reported excessive weight gain in exclusively breastfed infants during the first year when the mother's milk had variability in its macronutrient composition[191,192]. Further studies may be needed to seek the association between mother's milk macronutrient status and infant development.

On the positive side, the variability in concentration of mother's milk nutrients may ensure the health and protection of infants. The fatty acid composition of mother's milk changes in response to infections in mothers and/or infants[193]. Although the exact mechanisms of this variability, as well as the possible consequences, are unknown, it has been suggested that this variability could be physiologically significant as the fatty acids have antiviral and antibacterial properties [193]. It is also noteworthy that a significant increase in protein concentration of mother's milk is observed during the weaning period[194]. This increased protein content makes the milk produced during the late stages of lactation more suitable to preemies[194]. The milk produced during this stage also has a high content of antibacterial proteins such as lactoferrin and immunoglobulins[195].

Many anti-HIV factors including proteins lactoferrin[196], MUC1[197], and bile salt-stimulated lipase[198] are present in mother's milk of HIV-infected mothers. The risk of HIV transmission during breastfeeding is very low[199]. Moreover, milk from HIV-infected mothers may prevent transmission of HIV via multiple routes[200]. Considering breastfeeding benefits, the WHO also recommends HIV-infected mothers to exclusively breastfeed their infants in the first six months, in synergy with antiretroviral therapy[201].

**Ways of storing expressed milk**

Freshly expressed mother's milk is stored cold, more often frozen to be used after a long time to prevent the growth of microorganisms. There have been many studies regarding the optimal temperature for storage and the impact of storage on the nutrient composition of mother's milk. Many studies suggest that nutrient composition and quality of mother's milk may be affected by freezing, thawing, light exposure, pasteurization and storage container[202,203].

Studies that attempted to measure fat and/or caloric contents in frozen mother's milk have reported inconsistent findings which may be due to the difference in methods used in their study[204–206]. To measure fat and caloric contents in mother's milk, creamatocrit and human milk analyzer (HMA) are widely used. Creamatocrit involves adding few drops of mother's milk in a capillary tube and centrifuging them to separate the fat (*cream*) from the aqueous components of mother's milk. The ratio (%) of the length of the cream column over the length of the total milk column indicates the fat contents[207]. In HMA, a thin layer of mother's milk is exposed to infrared radiation and the calculations are done based the amount of radiations absorbed by different functional groups at specific wavelengths[206]. In creamatocrit, fat is the only measured variable. For example, to account for the effects of freezing on fat contents in mother's milk, Wang *et al.* used the following regression equations to calculate caloric contents in fresh and frozen mother's milk[207]: For fresh mother's milk, energy (kcal/dl) = 5.99 x creamatocrit(%) + 32.5; for frozen mother's milk, energy (kcal/dl) = 6.20 x creamatocrit(%) + 35.1[207]. On the other hand, in HMA, protein, lactose and fat in mother's milk can be measured[208]. For example, in a study that used HMA[206], caloric contents in mother's milk was calculated using this equation: energy (kcal/dl) = 9.25 x fat + 4.40 x total nitrogen + 3.95 x lactose[206]. In several studies it was observed that creamatocrit correlated strongly with fat and energy content in mother's milk[207,209,210]. Wang et al. observed that freezing and thawing of mother's milk did not alter this correlation[207]. They suggested that pasteurization may affect the accuracy of results[207,211]. However, Neill et al.

reported that creamatocrit overestimates fat and energy content in mother's milk and HMA has increased accuracy(208).

Mother's milk is often frozen at less than –20°C in neonatal units and human milk banks, while a better option is to freeze mother's milk at less than –80°C but the cost is much higher(206). A study on frozen (–20°C) mother's milk samples for 28 days, that used the creamatocrit method, reported that caloric contents did not change much(205). In a study that used the HMA method, Garcia-Lara et al. observed that frozen (–20°C) mother's milk contains less fat and calories than fresh mother's milk(206). They observed that decrease in the fat and caloric contents increase as a function of duration (7, 15, 30, 60, and 90 days) of freezing at less than –20°C, i.e. the longer the freezing duration, frozen mother's milk loses more fat and caloric contents(206). In their study, fresh mother's milk had 4.52-5.64 g/100 ml of fat and 0.72-0.82 kcal/ml of calories, whereas frozen mother's milk had only 3.84-5.14 g/100 ml of fat and 0.65-0.77 kcal/ml of calories after 90 days of freezing(206). The authors also note that different factors including the structural modification of free and glycerol-bound fatty acids due to lipid peroxidation (i.e oxidative degradation of lipids), freezing and thawing effect, and homogenization procedure could have biased their readings. In their study freezing for 7, 15, 30, 60, and 90 days at less than –20°C, did not significantly change the nitrogen content in frozen mother's milk with respect to fresh mother's milk, and the decrease of lactose in frozen mother's milk was only significant after 90 days of freezing(206).

Lactose concentration in mother's milk is also not affected by pasteurization and thawing procedures(203,212) while pasteurization can affect fat and protein concentration in mother's milk(213). Therefore, while delivering mother's milk to preterm infants, special care should be taken since the reduction in fat and protein concentration due to repeated processes preceding its delivery may affect their growth rate(213,214). Vieira et al. observed significant reduction in fat concentration delivered via continuous infusion feeding(213). This loss occurs due to the separation of fat from the aqueous fraction of mother's milk and its adherence to the delivery system(214). Solutions to prevent fat loss can be considered. For example, Jarjour et al. have shown that mixing mother's milk during continuous infusion can reduce this fat loss(214).

Freeze thawing procedures can break the milk fat globule membranes (MFGMs) in mother's milk resulting in coalescence(206,215). At –20°C, enzymes present in mother's milk may still be active(206). The triglycerides released via rupturing of MFGMs can come in contact with lipase enzymes(215). At –20°C and above, lipase catalyzes the hydrolysis of triglycerides (also known as lipolysis) into diglycerides, monoglycerides, and free fatty acids[216]. Triglycerides are esters derived from glycerol and three fatty acids. In a diglyceride, its glycerol is ester-linked to two fatty acids, whereas in a monoglyceride, its glycerol is ester-linked to only a fatty acid. However, at –70°C, fat hydrolysis does not occur[217]. Lipolysis also takes place at room temperature after 24 hours. Hamosh et al. observed lipolysis in mother's milk after 24 hours at 15°C, 25°C and 38°C[202]. Studies suggest that fatty acids produced by lipolysis have anti-microbial activity[218–220]. However, rupturing of MFGMs may reduce the bactericidal properties of mother's milk as MFGM has bacterial sequestration abilities[215]. Ogundele et al. observed that in addition to freezing, refrigeration and heat treatment also reduced the bactericidal activity of mother's milk[215]. They suggested that refrigeration (4°C) is preferable over freezing for short term storage, while for long term storage freezing (–20°C) can be used since they observed that after 1 month of freezing two-thirds of the bactericidal activity in mother's milk remained.

Lipid peroxidation is the process in which free radicals (most commonly reactive oxygen species) remove electrons from lipids producing reactive intermediates[221]. Turoli et al. observed lipid peroxidation in fresh mother's milk, formula milk and frozen (–20°C) mother's milk[222]. They observed highest lipid peroxidation in frozen mother's milk. This finding may be related to the higher contents of free fatty acids in frozen mother's milk due to lipolysis activity of lipase. Lipid peroxidation is also observed at temperatures of 4°C (refrigerating temperature)[223] and –80°C (freezing temperature)[224]. However, lipid peroxidation is minimal in frozen mother's milk (–20°C and –80°C) compared to that stored in refrigerator (4°C)[223,224]. Silvestre et al. suggested that if mother's milk is to be preserved over 15 days, then the freezing temperature of –80°C is preferable over –20°C[224] as it can minimize lipid peroxidation. Presence of lipid peroxidation products in infant feeds may play roles in development of diseases including necrotizing enterocolitis and bronchopulmonary dysplasia[225]. Studies to determine the effect of consuming lipid peroxides in premature babies, newborns, and infants are needed.

Even though the lipid peroxidation products in mother's milk poses serious health hazards to infants, the high antioxidant activity of mother's milk can protect the infants against the diseases. However, several studies that measured the antioxidant activity of mother's milk reported that it decreases with storage both at refrigeration (4°C) and freezing temperatures (–8°C, –20°C, –80°C)[224,226,227]. However, the complete list of antioxidant components in mother's milk and their contribution to this variation is still unknown. In a study, it was reported that carotenoid (major antioxidants in mother's milk) contents in frozen mother's milk kept at –18°C for 28 days were not significantly decreased(204). It is recommended that mother's milk should be stored at –80°C for a period of less than 30 days to maximally preserve antioxidant activity(224).

Freezing and thawing procedures can also cause protein denaturation(203,212). Freezing mother's milk can change and destabilize casein micelles and change the quaternary structure of whey proteins, resulting in the formation of precipitates in mother's milk(206). However, storage temperatures have not shown to influence the protein concentration in mother's milk(212).

Heat treatment or storage at –20°C/–70°C for longer duration have little effects on concentration of vitamins A, D and E in mother's milk(226,228,229) while levels of vitamins C and B6 may get affected. According to Buss et al. storage for more than one month in a freezer (-16°C) or 24 hours in a refrigerator (4°C-6°C) may substantially reduce vitamin C concentration in mother's milk(230). Holder pasteurization can affect the vitamin B6 concentration in mother's milk(228) .

Many protective immunologic components of frozen mother's milk kept at –20°C remain stable for a duration of up to 365 days while storage container, freezing and/or thawing have shown to affect lactoferrin and live cells level(231)). However, lactoferrin levels in mother's milk significantly decreases after 3 months of freezing (-20°C)(232). Cycles of freezing and thawing can rupture cells in mother's milk(233) and longer storage time reduces the cell function(231). Frozen mother's milk kept at –20°C has lower number of cells and functions of surviving cells than fresh mother's milk does(231). Cells in mother's milk can also adhere to the surface of the storage container particularly Pyrex glass container(231,234) resulting in their loss.

Good mother's milk storage practices should be adopted to ensure the health and safety of infants. For example, It is known that containers made with bisphenol A (BPA) can adversely affect the infant, therefore such containers should be avoided for mother's milk storage(235,236). Eglash et al. suggest that usage of bisphenol S, a BPA alternative, in containers may also harm the baby(236). Frozen mother's milk should be kept away from the freezer door opening, and from the walls of the freezer if it does self defrosting(236). Adding fresh mother's milk to cooled milk should also be avoided[236]. More details on optimal mother's milk storage practices can be found elsewhere[236].

**Duration of mother's milk intake**

The WHO recommends *exclusive* breastfeeding for the first six months of the infant and continued breastfeeding up to 2 years of age or beyond along with complementary foods[1]. In support of the WHO recommendation, several studies suggested that benefits of breastfeeding are dose dependent and exclusive breastfeeding for the first six months provides optimal nutrition and health protection to the infant[237,238]. On the other hand, some studies suggested that exclusive breastfeeding may not provide sufficient energy to a 6 month old infant[239,240]. However, there is no doubt that breastfeeding can provide substantial amount of essential nutrients including proteins, fats and most vitamins well beyond 1 year[238,241].

Exclusivity and longer duration of mother's milk intake can influence its benefits. A study that compared exclusive mother's milk intake for 6 months or more and that for 4 to less than 6 months in the United States, reported that the former has a lower risk for respiratory tract infections, including pneumonia and recurrent otitis media (inflammatory diseases of the middle ear)[233]. The risk for pneumonia was reported to be 1.6% for exclusive mother's milk intake for 6 months, and 6.5% for exclusive mother's milk intake for 4 to less than 6 months[233]. Breastfeeding for longer duration (at least 6 months) can provide better protection against respiratory and gastrointestinal tract infections[24,25,31]. Doses of mother's milk are positively associated with growth rate; infants who consumed more mother's milk, have more lean body mass and exclusively breastfed infants are taller than mixed-fed infants[242].

However, there are inconsistencies in recommendations for the duration for exclusive breastfeeding. In a study, Kramer et al. compared exclusive breastfeeding for 6 months vs exclusive breastfeeding for 3 to 4 months with continued partial breastfeeding until 6 months[243]. Infants in the formal category experienced less morbidities from gastrointestinal tract infections compared to those in the latter group[243]. They didn’t observe any risks in recommending exclusive breastfeeding for the first six months of life. There were also no growth deficits in infants who are exclusively breastfed for 6 months. Moreover, mothers of such infants experienced more rapid weight loss and delayed return of menstrual periods[243]. On the other hand, existing studies suggest that milk transfer from mother to infant do not increase significantly with age[239]. Reilly et al. observed that mother's milk metabolizable energy content is probably lower than usually assumed[239]. They hypothesized that milk produced by many mothers may not be sufficient to feed infants at 6 months and suggested that more research is needed on the energy intake of exclusively breastfed infants at 5-6 months[239,240]. Previous studies also suggest that the scientific evidence to support the recommendation for exclusive breastfeeding until 6 months is weak and large high quality randomized studies are needed to identify potential benefits and risks of exclusive breastfeeding for 6 months[240,243]. However, early introduction of complementary foods may negatively affect the infant[244–246]. WHO also suggests that complementary foods are less likely to be effective in providing limiting nutrients to infants as compared to medicinal supplementation before 6 months[247]. WHO recommends medicinal iron drops for low birth weight infants and infants of mothers with poor prenatal iron status, before 6 months, since they are at high risks for iron deficiency[247].

In agreement with the WHO recommendation of continued breastfeeding until 2 years or beyond, existing studies suggest that prolonged breastfeeding after 6 months along with complementary foods is beneficial for infants. For example, recently, Weaver et al. observed that longer durations of breastfeeding (assessed upto 3 years) can predict maternal sensitivity to her child up to age of 11 years [248)]. In another study, Perrine et al. observed that longer durations (assessed up to ≥1 year ) and exclusivity (assessed up to ≥3 months) of breastfeeding are positively associated with healthy dietary patterns at 6 years of age of children[249]. Vogazianos *et al.* observed the protective effect of breastfeeding against acute otitis media for the first 11 months[250]. In their study, even though continuation of breastfeeding up to the 18 months has shown some protective effect, the results were not statistically significant. A longer period of mother's milk consumption also seems to be related to a decreased risk of asthma[33]

Continuation of breastfeeding and introduction of complementary foods are both essential after 6 months of age of the infant. For example, it is known that the levels of certain nutrients in mother's milk including zinc, calcium, vitamin B6 and vitamin C decreases in the course of lactation[251,252], which emphasizes the introduction of complementary foods after 6 months. On the other hand, mother's milk is an irreplaceable source of vitamin A in the first three years[241,253] and an important source of key nutrients including fat in the first two years of the infant's life[241,254]. Therefore, early cessation of breastfeeding can make it difficult for the child to receive certain nutrients that are poorly available from other foods. For example, in a study that assessed contribution of mother's milk to toddler diets in western Kenya, Onyango et al. observed that upon termination of breastfeeding before 2 years, the amount of food intake by the infant increased quantitatively, but it did not compensate for the quality of mother's milk[(241)]. Additionally, the immunological protection from mother's milk extends even after 1 year and may persist at least until the cessation of breastfeeding. For example, in a longitudinal study of mother's milk composition, it was found that mother's milk in the second year of lactation had significantly higher levels of protein, lactoferrin, immunoglobulin A and lysozyme than milk bank samples from donors less than one year postpartum[(255)].

Frequent and on demand breastfeeding is recommended by WHO until 2 years or beyond[(247)]. Energy content in mother's milk is about 0.7 kcal/ml[(256)]. In developing countries average energy intake from mother's milk is estimated to be 423 kcal/day at 6-8 months, 379 kcal/day at 9-11 months and 346 kcal/day at 12-23 months of age[(247)]. Healthy breastfed infants require total energy intake of 615 kcal/day at 6-8 months, 686 kcal/day at 9-11 months and 894 kcal/day at 12-23 months of age[(247,257)]. By subtracting the energy intake provided by mother's milk from total energy, amount of energy needed from complementary foods can evaluated. The importance of solid foods on infant nutrition in later ages cannot be underestimated. In an article published by Canadian Pediatric Society[(258)], the authors state that "After 12 months of age, your baby should not take more than 24 ounces (~= 710 ml) of milk products per day. Otherwise, she'll fill up and won't want to eat solid foods. Also, she may develop iron deficiency anemia."

Despite having numerous benefits, there are several barriers in implementing breastfeeding according to WHO recommendations and breastfeeding rates decreases with duration in many areas. In a study of breastfeeding duration among the population of Perth, Australia, only 45.9% of babies were receiving any mother's milk and only 12% were exclusively receiving full mother's milk at 6 months of age[(259)]. At 12 months of age, only 19.2% of babies were receiving any mother's milk. Factors that positively influence the duration of mother's milk intake include positive attitudes of mothers[(259)]. Factors that negatively influence the duration of mother's milk intake include early breastfeeding difficulties (especially in the first 4 weeks)[(259)], inadequate mother's milk[(260)], lack of support from family [261)], introduction of a pacifier before 10 weeks of age[(259)], smoking[(259)], social stigma[(262)] and short maternity leave/early return to work[259]. Studies show that lactation counseling can promote breastfeeding exclusivity and its total duration[263,264].

**Relactation**

Relactation refers to the process of (i) re-establishing a supply of mother's milk that has reduced or ceased[265] and/or (ii) re-initiating breastfeeding. Women who have stopped breastfeeding recently or in the past can successfully relactate even without a further pregnancy and produce enough milk to exclusively breastfeed her child[266]. When women receive necessary support from the community and health care services, relactation is rarely necessary[266]. However, in certain occasions such as separation of mother and infant due to weak health conditions of the mother or infant, relactation may be necessary. Physiology-based methods and galactagogues are usually used for relactation. For the success of these methods, support and motivation from family, public and health care providers are also necessary for women.

For successful relactation, maximum stimulation of the breast and nipples especially through infant suckling the breast is recommended[266]. The physiological basis of for this recommendation is that nipple stimulation especially through suckling promotes the release of two hormones from pituitary gland in brain- prolactin and oxytocin[267]. Prolactin stimulates the growth of secretory alveoli in breasts and the production of milk from the cells of the alveoli[266]. Oxytocin helps in pressing out milk from the breasts by causing muscle cells surrounding the alveoli to contract[266]. Oxytocin production is also affected by mother's mental state[267]. Hence ensuring mental wellbeing of mothers is also necessary to support this process. To encourage infants suckle the breasts for a long time, breastfeeding supplements[266,268] or 'drop and drip' method[269] are usually used.

When the physiology-based methods alone are not effective, galactagogues (or lactogogues) are considered[266]. Galactagogues are agents that are believed to promote the secretion and flow of milk. Examples of galactagogues include herbs such as fenugreek and prescription drugs such as domperidone and metoclopramide[270] (these drugs have side effects[270,271] and should only be used under medical supervision). Galactogogues are only effective in conjunction with full stimulation of the breasts[270]. Studies report inconsistencies regarding the effectiveness of galactagogues. For example, Seema et al. studied 50 mothers of infants less than 4 months old with complete and partial lactation failure, and observed that nipple stimulation by repeated suckling, along with motivation and support alone, helped 49 mothers to successfully relactate[272]. 46 mothers achieved complete relactation. In their study, employment of metoclopramide (galactagogue) did not have any significant positive effect on relactation. However, in another study involving 80 mothers who were expressing less milk for their infants in neonatal intensive care unit, the usage of domperidone and metoclopramide increased the production of milk[271]. These results suggest that the effects of galactagogues on mother's milk production require further investigation.

Mothers need to be patient while trying to relactate since it may take more than a month to completely relactate. In the study of Seema et al.[272], it took 2-6 days for the first mother's milk secretion to appear and 7-60 days for complete relactation. The process of relactation is also challenging and mothers need to have great will power, determination and ability to access support for its success[265]. A study involving 10 participants who attempted relactation[265] highlighted that challenges, such as colic and latching issues, could have impacted the initial breastfeeding and relactation experience and those having colicky babies reported that they didn't receive proper support from pediatricians and health care services regarding how to deal with colic. The participants also experienced common feelings including stress, rejection, and anger. With the numerous benefits associated with breastfeeding, it is important to create awareness among the public about the necessity to support breastfeeding and equally important is to educate the public to not ostracize mothers who cannot breastfeed[265].